\documentclass[12pt,preprint]{aastex}

\usepackage{amsmath}
\usepackage{appendix}
\usepackage{subfigure}
\usepackage{booktabs}

\begin{document}

\title{ON THE FORMATION OF PSR J1640+2224: A NEUTRON STAR BORN MASSIVE?
}
\author{
Zhu-Ling Deng$^{1,2,3,4,5}$, Zhi-Fu Gao$^{1,2,3}$, Xiang-Dong Li$^{4,5}$,  and Yong Shao$^{4,5}$}

\affil{$^{1}$Xinjiang Astronomical Observatory, Chinese Academy of Sciences, 150, Science 1-Street, Urumqi, Xinjiang 830011, China}

\affil{$^{2}$Key Laboratory of Radio Astronomy, Chinese Academy of Sciences, Nanjing 210008, China}

\affil{$^{3}$University of Chinese Academy of Sciences, 19A Yuquan Road, Beijing 100049, China; zhifugao@xao.ac.cn}

\affil{$^{4}$School of Astronomy and Space Science, Nanjing University, Nanjing 210023, China; lixd@nju.edu.cn}

\affil{$^{5}$Key Laboratory of Modern Astronomy and Astrophysics (Nanjing University), Ministry of
Education, Nanjing 210023, China}

\begin{abstract}
PSR J1640+2224 is a binary millisecond pulsar (BMSP)  with a white dwarf (WD) companion.  Recent observations indicate that the WD is very likely to be a $\sim 0.7\,M_{\odot}$ CO WD. Thus the BMSP should have evolved from an intermediate-mass X-ray binary (IMXB). However, previous investigations on IMXB evolution predict that the orbital periods of the resultant BMSPs are generally $<40$ days, in contrast with the 175 day orbital period of PSR J1640+2224. In this paper, we explore the influence of the mass of the neutron star (NS) and the chemical compositions of the companion star on the formation of BMSPs. Our results show that, the final orbital period becomes longer with increasing NS mass, and the WD mass becomes larger with decreasing metallicity. In particular, to reproduce the properties of PSR J1640+2224, the NS was likely born massive ($>2.0\,M_{\odot}$).

\end{abstract}


\section{INTRODUCTION}


Millisecond pulsars (MSPs) are neutron stars (NSs) with short spin periods ($P\leq 30$ ms) and weak surface magnetic fields \citep[$B \sim 10^{8}-10^{9}$ G;][]{L08}. They are believed to be old NSs that have been reactivated by accretion from their companion stars during the low-mass X-ray binary (LMXB) phase \citep{R82,A82}. The mass transferred from the companion star not only causes the decay of the NS's magnetic field, but also accelerates the NS's spin period to milliseconds \citep[see, e.g.,][for reviews]{B91,T06}. This recycling scenario has been strongly supported by the discovery of several X-ray pulsars with millisecond periods in LMXBs \citep[see][for a summary]{P2012}, and the discovery of the transition between a rotation-powered MSP state and a LMXB state in PSR J1023+0038 \citep{A09}, IGR J18245-2452 \citep{P2013}, and PSR J1227-4853 \citep{R15}.


When the mass transfer ceases, some of the LMXBs evolve to be binary systems consisting of a MSP and a He white dwarf (WD). This population is called low-mass binary pulsars (LMBPs) \citep{T06}. Compared with LMBPs, there is another population called intermediate-mass binary pulsars (IMBPs), which contain a pulsar with a spin period of tens of milliseconds and a CO or ONeMg WD with mass $\geq 0.4M_{\odot}$ \citep{C96,C01,C11}.
Most IMBPs are thought to have evolved from intermediate-mass X-ray binaries (IMXBs). Because of the relatively high mass ratio ($q=M_{2}/M_{1}>1.5$, where $M_{1}$ and $M_{2}$ are the NS mass and the donor star mass, respectively), it had been suggested that the mass transfer could be unstable, and the NS would spiral into the envelope of the donor in less than a millennium \citep{P76,W84,I93}. More recent studies showed that X-ray binary systems with an intermediate-mass donor star may avoid entering the spiral-in phase, undergoing rapid mass transfer on a (sub)thermal timescale, and eventually evolve into IMBPs \citep[e.g.,][]{T00,P00,K2000,S12}. Since the birth rate of IMXBs is significantly larger than that of original LMXBs, and since IMXBs likely evolve to resemble observed LMXBs when the mass ratio becomes less than unity, it is generally believed that the majority of LMXBs may have started their lives with an intermediate-mass donor star \citep[e.g.,][]{P03}.

PSR J1640+2224 is a MSP with a spin period $P=3.16$ ms. It is located in a wide, nearly circular binary system (orbital period $P_{\rm orb}=175$ days and orbital eccentricity $e=7.9725\times 10^{-4}$) with a WD companion \citep{L96}. These characteristics seem to indicate that this system has evolved from a wide LMXB \citep{T11b}. According to the theoretical correlation between the orbital period and the WD mass \citep{R95,T99,L11,J14}, the WD companion's mass would be $\sim (0.35-0.39)\,M_{\odot}$ depending on the metallicity. \citet{V18} recently presented the first astrometric parallax measurement of PSR J1640+2224, based on the observations taken with the Very Long Baseline Array. Using the new distance in the analysis of the Hubble Space Telescope observation, they found that the WD mass is $0.71^{+0.21}_{-0.20}\,M_{\odot}$ and $0.66^{+0.21}_{-0.19}\,M_{\odot}$ for DA and DB WDs, respectively. The results indicate that the WD mass is larger than $0.4M_{\odot}$ with $>90\%$-confidence, so it is most likely a CO WD. However, previous works predicted that most IMBPs formed from IMXBs have a spin period of tens of milliseconds and an orbital period $P_{\rm orb}<40$ days \citep{T00,P00,P02,P03,S12}. These characteristics are in conflict with the observational parameters of PSR J1640+2224. Therefore, the formation of this system had been a puzzle.

In this paper, we try to explore the formation route of PSR J1640+2224, paying particular attention to the influence of the NS mass and the metallicity of the donor star. The reminder of this  paper is organized as follows. We describe the stellar evolution code and the binary model in Section 2. The calculated binary evolution results are demonstrated in Section 3. We present our discussion and summarize in Section 4.

\section{BINARY EVOLUTIONARY CALCULATIONS}

\subsection{The stellar evolution code}
All the calculations were carried out by using the stellar evolution code Modules for Experiments in Stellar Astrophysics (MESA; version number 11554; Paxton et al. 2011, 2013, 2015, 2018, 2019). We have calculated the evolutions of a large number of I/LMXBs, adopting both Population \uppercase\expandafter{\romannumeral1} ($X=0.7$ and $Z=0.02$) and \uppercase\expandafter{\romannumeral2} ($X=0.75$ and $Z=0.001$) chemical compositions for the donor stars. For the treatment of convection in the donor stars, we employ both exponential diffusive overshooting with the parameter $f_{\rm ov}=0.01-0.016$ \citep{H00} and a mixing-length parameter of $\alpha=2.0$ (see Section 4). In our calculations, we have considered a number of binary interactions to follow the details of mass transfer processes, including gravitational radiation (GR), magnetic braking (MB) and mass loss, which lead to orbital angular momentum loss.

\subsection{The input physics}
We assume that the binary initially consists of a NS of mass $M_{1}$ and a zero-age main-sequence donor of mass $M_{2}$. The Roche-lobe (RL) radius $R_{\rm L,2}$ of the donor is evaluated with the formula proposed by \citet{E83},
\begin{equation}
\frac{R_{\rm L,2}}{a}=\frac{0.49 q^{-2 / 3}}{0.6 q^{-2 / 3}+\ln \left(1+q^{-1 / 3}\right)},
\end{equation}
where $a$ is the orbital separation of the binary and $q=M_{2}/M_{1}$ is the mass ratio. We adopt the \citet{R88} scheme to calculate the mass transfer rate via Roche-lobe overflow (RLOF),
\begin{equation}
-\dot{M}_{2}=\dot{M}_{2,0} \exp \left(-\frac{R_{2}-R_{\mathrm{L}, 2}}{H}\right),
\end{equation}
where $H$ is the scale-height of the atmosphere evaluated at the surface of the donor, $R_{2}$ is the radius of the donor, and
\begin{equation}
\dot{M}_{2,0}=\frac{1}{e^{1 / 2}} \rho c_{\mathrm{th}} Q,
\end{equation}
where $\rho$ and $c_{\mathrm{th}}$ are the mass density and the sound speed on the surface of the star respectively, and $Q$ is the cross section of the mass flow via the $L_{1}$ point. The very small orbital eccentricity indicates that tides keep the binary orbit circular \citep{K88}, so the orbital angular momentum is
\begin{equation}
J_{\mathrm{orb}}=\frac{M_{1} M_{2}}{M_{1}+M_{2}} \Omega a^{2},
\end{equation}
with the orbital angular velocity $\Omega=\sqrt{GM/a^{3}}$ (where $M=M_1+M_2$ is the total mass). Taking the logarithmic derivative of Eq.~(4) with respect to time gives the rate of change in the orbital separation
\begin{equation}
\frac{\dot{a}}{a}=2 \frac{\dot{J}_{\mathrm{orb}}}{J_{\mathrm{orb}}}-2 \frac{\dot{M}_{1}}{M_{1}}-2 \frac{\dot{M}_{2}}{M_{2}}+\frac{\dot{M}}{M}.
\end{equation}
Here the total rate of change in the orbital angular momentum is determined by
\begin{equation}
\frac{\dot{J}_{\mathrm{orb}}}{J_{\mathrm{orb}}}=\frac{\dot{J}_{\mathrm{gr}}}{J_{\mathrm{orb}}}+\frac{\dot{J}_{\mathrm{mb}}}{J_{\mathrm{orb}}}+\frac{\dot{J}_{\mathrm{ml}}}{J_{\mathrm{orb}}}.
\end{equation}
The three terms on the right-hand-side of Eq.~(6) represent angular momentum losses caused by GR, MB, and mass loss, respectively. The GR-induced rate $\dot{J}_{\rm gr}$ is calculated with the standard formula \citep{L59,F71}
\begin{equation}
\frac{\dot{J}_{\mathrm{gr}}}{J_{\mathrm{orb}}}=-\frac{32 G^{3}}{5 c^{5}} \frac{M_{1} M_{2} M}{a^{4}},
\end{equation}
where $G$ and $c$ are the gravitational constant and the speed of light, respectively. The prescription of \citet{V81} is adopted to calculate the angular momentum loss due to MB,
\begin{equation}
\frac{\dot{J}_{\mathrm{mb}}}{J_{\mathrm{orb}}}=-3.8 \times 10^{-30} \frac{G R_{2}^{4} M^{2}}{a^{5} M_{1}} \mathrm{s}^{-1}.
\end{equation}
For IMXBs and wide LMXBs, the mass transfer rate $|\dot{M_{2}}|$ may be higher than the Eddington-limit accretion rate $\dot{M}_{\mathrm{Edd}}$ of the NS. Thus, the accretion rate of the NS is evaluated using the following formula:
\begin{equation}
\dot{M}_{\mathrm{1}}=\min \left(\left|\dot{M}_{2}\right|, \dot{M}_{\mathrm{Edd}}\right),
\end{equation}
and the mass loss rate from the binary system is:
\begin{equation}
\dot{M}=\dot{M}_{1}-\left|\dot{M}_{2}\right|.
\end{equation}
In the case of super-Eddington accretion, we adopt the isotropic reemission model, assuming that the extra material leaves the binary in the form of isotropic wind from the NS. Therefore, the angular momentum loss rate due to mass loss can be derived to be
\begin{equation}
\dot{J}_{\mathrm{ml}}=-\left(\left|\dot{M}_{2}\right|-\dot{M}_{\mathrm{1}}\right) a_{\mathrm{1}}^{2} \Omega,
\end{equation}
where $a_{\mathrm{1}}$ is the distance between the NS and the center of mass of the binary.

\section{RESULTS OF EVOLUTION CALCULATIONS}

We have performed calculations of the binary evolution for thousands of I/LMXB systems. We choose the initial NS mass $1.4\,M_{\odot}\leq M_{1}\leq 2.2\,M_{\odot}$ and the initial donor mass $1.0\,M_{\odot}\leq M_{2}\leq 4.0\,M_{\odot}$, and set the initial orbital period 1.0 day $\leq P_{\rm orb}\leq 60$ days. We use exponential diffusive overshooting for the donor stars.  Note that there exists a bifurcation period for the initial orbital period $P_{\rm bf}\sim 1$ day \citep{Py88,Py89}. If the initial orbital period is below $P_{\rm bf}$, the binary systems will evolve with shrinking orbits, possibly forming ultra-compact binaries. This kind of evolution cannot reproduce PSR J1640+2224, and will not be considered here. In addition, we exclude the evolutions with the final orbital periods exceeding 500 days.

\begin{figure}

\centerline{\includegraphics[scale=0.6]{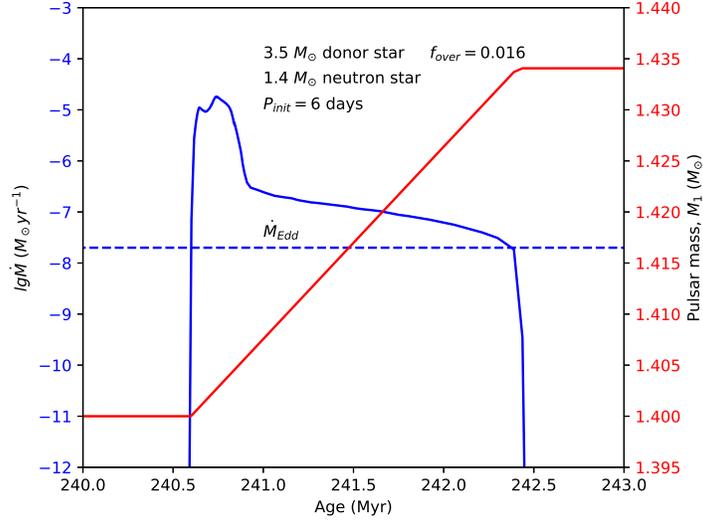}}
\caption{Example evolution of a typical IMXB consisting of a NS and a donor star  with initial masses $M_{1}=1.4M_{\odot}$ and $M_{2}=3.5M_{\odot}$ respectively, and orbital period $P_{\rm orb}=6$ days. The blue and red solid curves denote the evolutionary tracks for the mass transfer rate and the NS mass, respectively, and the blue dashed curve the Eddington-limit accretion rate.
   \label{figure1}}

\end{figure}

\begin{figure}

\centerline{\includegraphics[scale=0.6]{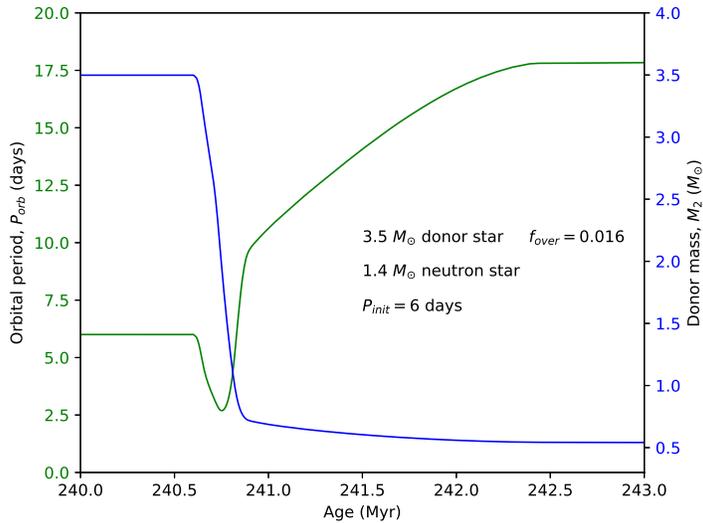}}
\caption{Same as Fig.~1. The green and blue curves denote the evolutionary tracks for the orbital period and the donor mass, respectively.
   \label{figure2}}

\end{figure}

\subsection{A typical example for the IMXB evolution}
We present the evolutionary sequences for a typical IMXB system that consists of a $1.4M_{\odot}$ NS and a $3.5M_{\odot}$ donor star with an initial orbital period $P_{\rm orb}=6$ days in Figs.~1 and 2. The donor star starts to fill its RL and commence mass transfer at the age of 240.6 Myr. The mass transfer process lasts about 1.9 Myr, at a rate higher than the Eddington accretion rate (the maximum mass transfer rate $>$ $10^{-5}M_{\odot}$\, yr$^{-1}$). In this case, most of the transferred material from the donor is lost from the binary system. Thus, after the mass transfer, the donor evolves into a CO WD of mass $0.543M_{\odot}$, while the NS has accreted only $0.034M_{\odot}$ mass. The orbital period becomes smaller in the former stage of the mass transfer process due to the relatively high mass ratio ($>$1); when the mass ratio is reversed, the orbit begins to expand. The final orbital period is $P_{\rm orb}=17.8$ days.

\begin{figure}

\centerline{\includegraphics[scale=0.6]{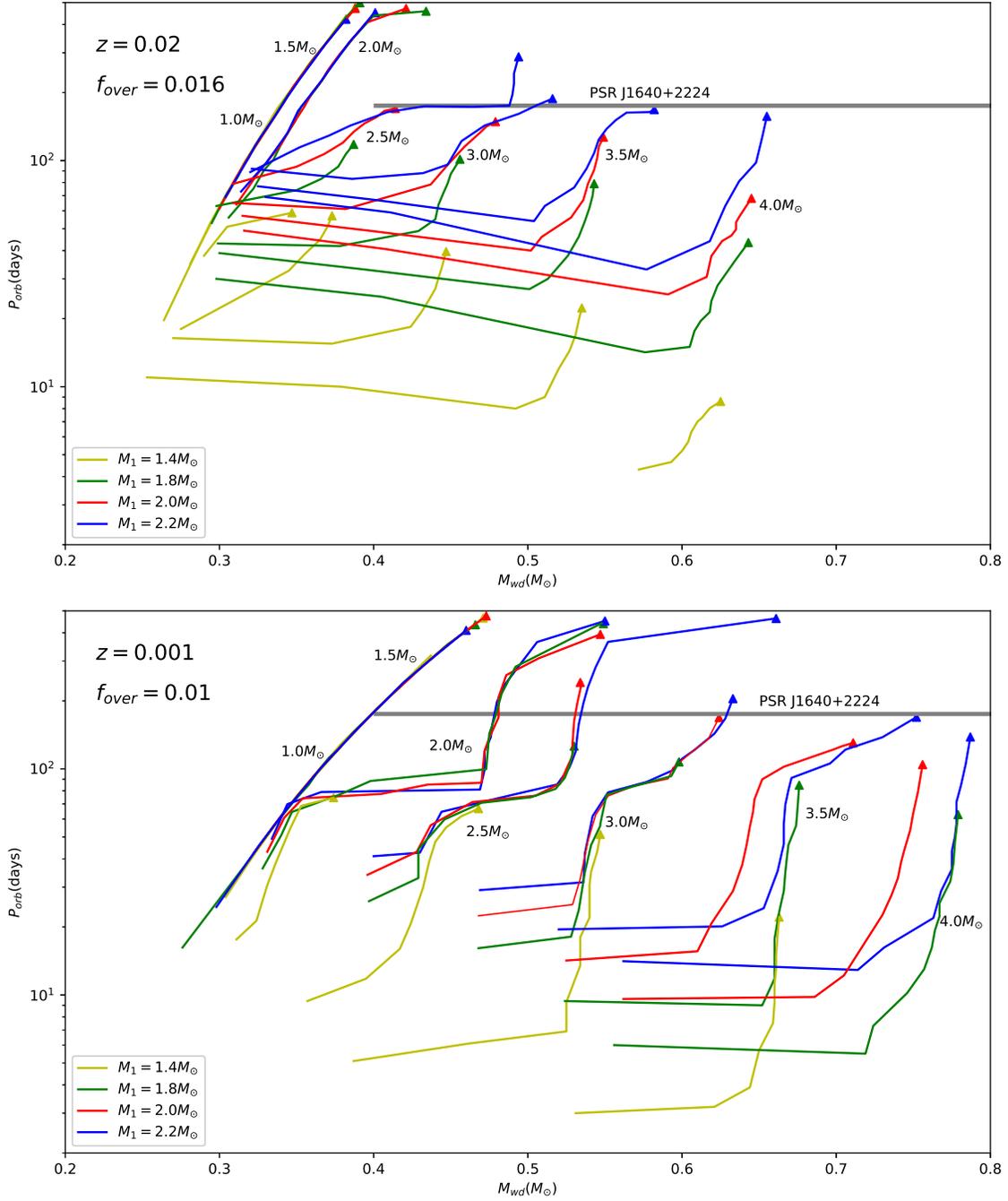}}
\caption{The final orbital period as a function of the WD mass. The yellow, green, red and blue curves are for the initial NS mass of $1.4M_{\odot}$,  $1.8M_{\odot}$, $2.0M_{\odot}$ and $2.2M_{\odot}$, respectively. Beside each curve we label the initial donor mass. The gray, solid horizontal line represents the distribution of the orbital period and the WD mass for PSR J1640+2224. The triangles are used to distinguish different curves that overlap. In the top and bottom panels we adopt different chemical compositions and overshooting parameters (top: Population \uppercase\expandafter{\romannumeral1} and $f_{\rm ov}=0.016$; bottom: Population \uppercase\expandafter{\romannumeral2} and $f_{\rm ov}=0.01$).
   \label{figure3}}

\end{figure}

\subsection{The $P_{\rm orb}^{\rm final}-M_{\rm WD}$ diagram}
Fig.~3 shows the calculated final orbital period as a function of the WD mass for different initial masses of the donor star and the NS. The yellow, green, red, and blue curves are for NSs with the initial mass of $1.4M_{\odot}$, $1.8M_{\odot}$, $2.0M_{\odot}$, and $2.2M_{\odot}$, respectively. Beside each curve we label the initial mass of the donor star. The triangles are used to denote the final systems with longest orbital periods for a give donor star, and also to distinguish different curves that  overlap. The gray horizontal line represents the orbital period $-$ WD mass distribution for PSR J1640+2224.


In the top panel we show the results with Population I chemical compositions. We first examine the orbital period - WD mass relation for a $1.4 M_{\odot}$ NS. In this case, it is clear to see that, for a given WD mass $> 0.4\,M_{\odot}$, the predicted final orbital period is significantly shorter than 175 days. For more massive donor star, the resultant WD is more massive, while the final orbital period is shorter. So they are obviously unable to account for the properties of PSR J1640+2224. If we increase the mass of the NS, the final orbital period becomes longer. When $M_{1}= 2.0\,M_{\odot}$, it is possible to simultaneously account for the WD of mass $\sim 0.4 \,M_{\odot}$ and the orbital period $\sim$ 175 days. Moreover, if the WD is more massive than $0.6\,M_{\odot}$, then the initial NS mass should be enhanced to be $2.2\,M_{\odot}$.

The bottom panel shows the results with Population II chemical compositions. It is known that lower metallicities lead to smaller stellar radius and shorter nuclear evolutionary timescale. So stars with the same mass but lower metallicities form WDs in a narrower orbit \citep{T99,J14}. We can see that the evolution of LMXBs containing a $1.4M_{\odot}$ NS and a $1.0M_{\odot}$ donor star may form PSR J1640+2224-like binaries with a $0.4M_{\odot}$ WD companion. However, if the WD indeed has a mass higher than $0.6M_{\odot}$, then we still require the NS to be initially more massive than $2.0M_{\odot}$.

We summarize our calculated results with $M_{\rm WD} \geq 0.4 M_{\odot}$ and the final orbital period 170 $\leq P_{\rm orb} \leq$ 180 days in Table 1. It is noted that PSR J1640+2224 is most likely to have descended from an IMXB via Case B RLOF. Tables 2-5 present more general results by taking into account different initial NS mass.

\begin{table}
\footnotesize
\begin{center}
\caption{Parameters for IMXB evolution with the final orbit period 170 days $\leq P_{\rm orb}\leq 180$ days and the WD mass $M_{\rm WD}\geq0.4\,M_{\odot}$}
\begin{tabular}{ccccccccccc}
  \toprule[2pt]
  $M^{\rm ini}_{1}$ & $M^{\rm ini}_{2}$ & $P^{\rm ini}_{\rm orb}$ & $t_{\rm RLO}$  & $M^{\rm fin}_{1}$ & $M^{\rm fin}_{2}$ & $P^{\rm fin}_{\rm orb}$ &  & $\bigtriangleup M_{1}$ & $\bigtriangleup t_{\dot{M}}$ & $\dot{M}_{\rm max}$   \\
  ($M_{\odot}$) & ($M_{\odot}$) &    (days)       & (Myr)        & ($M_{\odot}$) & ($M_{\odot}$) &     (days)      &   Type   &    ($M_{\odot}$)       &           (Myr)              & ($M_{\odot}$ yr$^{-1}$)   \\
  (1) & (2) & (3) & (4) & (5) & (6) & (7) & (8) & (9) & (10) & (11)    \\

  \hline
    \multicolumn{11}{c}{Population \uppercase\expandafter{\romannumeral1} ($X=0.7$, $Z=0.02$)} \\
  \hline
  $2.0$  & 2.5 & 5.35 & 597.900  & 2.109 & 0.414 & 170.0 & HeCO & 0.109 & 5.845 & $3.16\times 10^{-4}$    \\
  $2.2$  & 2.5 & 7.0 & 598.596  & 2.304 & 0.488 & 175.0 & CO & 0.104 & 6.580 & $5.01\times 10^{-6}$    \\
  $2.2$  & 3.0 & 7.0 & 363.239 & 2.255 & 0.506 & 176.5 & CO & 0.055 & 3.363 & $7.94\times 10^{-6}$    \\
  $2.2$  & 3.5 & 10.8 & 241.819  & 2.226 & 0.571 & 171.5 & CO & 0.026 & 1.558 & $2.51\times 10^{-5}$    \\
  $2.25$  & 4.0 & 15.2 & 170.965  & 2.266 & 0.634 & 170.8 & CO & 0.016 & 0.857 & $6.31\times 10^{-5}$    \\
  \hline
  \multicolumn{11}{c}{Population \uppercase\expandafter{\romannumeral2} ($X=0.75$, $Z=0.001$)} \\
  \hline
  $1.4$  & 1.0 & 29.0 & 5965.290  & 1.732 & 0.400 & 177.9 & He & 0.332 & 19.440 & $7.00\times 10^{-8}$    \\
  $1.4$  & 1.5 & 17.0 & 1557.870  & 1.837 & 0.400 & 173.9 & He & 0.437 & 27.950 & $3.16\times 10^{-7}$    \\
  $1.8$  & 1.5 & 12.5 & 1547.698  & 2.427 & 0.400 & 175.4 & He & 0.627 & 38.912 & $7.59\times 10^{-8}$    \\
  $1.8$  & 2.0 & 12.0 & 642.987  & 2.046 & 0.479 & 171.4 & CO & 0.246 & 13.620 & $1.58\times 10^{-6}$    \\
  $2.0$  & 1.5 & 11.5 & 1544.370  & 2.694 & 0.400 & 178.3 & He & 0.694 & 42.240 & $5.89\times 10^{-8}$    \\
  $2.0$  & 2.0 & 10.5 & 650.153  & 2.293 & 0.484 & 174.1 & CO & 0.293 & 16.379 & $5.01\times 10^{-7}$    \\
  $2.0$  & 2.5 & 10.5 & 370.271  & 2.110 & 0.525 & 174.3 & CO & 0.110 & 6.503 & $7.94\times 10^{-6}$    \\
  $2.2$  & 1.5 & 10.5 & 1540.127  & 2.965 & 0.400 & 176.9 & He & 0.756 & 46.617 & $4.37\times 10^{-8}$    \\
  $2.2$  & 2.0 & 9.0 & 648.640  & 2.524 & 0.484 & 170.7 & CO & 0.324 & 13.085 & $3.16\times 10^{-7}$    \\
  $2.2$  & 2.5 & 8.5 & 369.724  & 2.32 & 0.525 & 173.2 & CO & 0.120 & 7.045 & $6.31\times 10^{-6}$    \\
  $2.2$  & 3.0 & 12.3 & 236.243  & 2.262 & 0.616 & 178.8 & CO & 0.062 & 4.155 & $2.85\times 10^{-5}$    \\
  $2.2$  & 3.5 & 19.0 & 169.025  & 2.246 & 0.723 & 177.3 & CO & 0.046 & 3.098 & $7.94\times 10^{-5}$    \\
  $2.2$  & 4.0 & 25.5 & 125.904  & 2.231 & 0.794 & 170.0 & CO & 0.031 & 1.760 & $2.00\times 10^{-4}$    \\
  \bottomrule[2pt]

\end{tabular}

\end{center}
Note.---Col.~(1): the initial NS mass. Col.~(2): the initial donor mass. Col.~(3): the initial orbital period. Col.~(4): the age of the donor star at the onset of RLOF. Col.~(5): the final mass of the NS. Col.~(6): the final mass of the WD. Col.~(7): final orbital period. Col.~(8): the type of the WD. Col.~(9): the mass accreted by the NS. Col.~(10): the duration of the mass transfer. Col.~(11): the maximum mass transfer rate.
\end{table}

\subsection{Further constraint from spin evolution}

\begin{figure}
\centering
\subfigure{\includegraphics[scale=0.5]{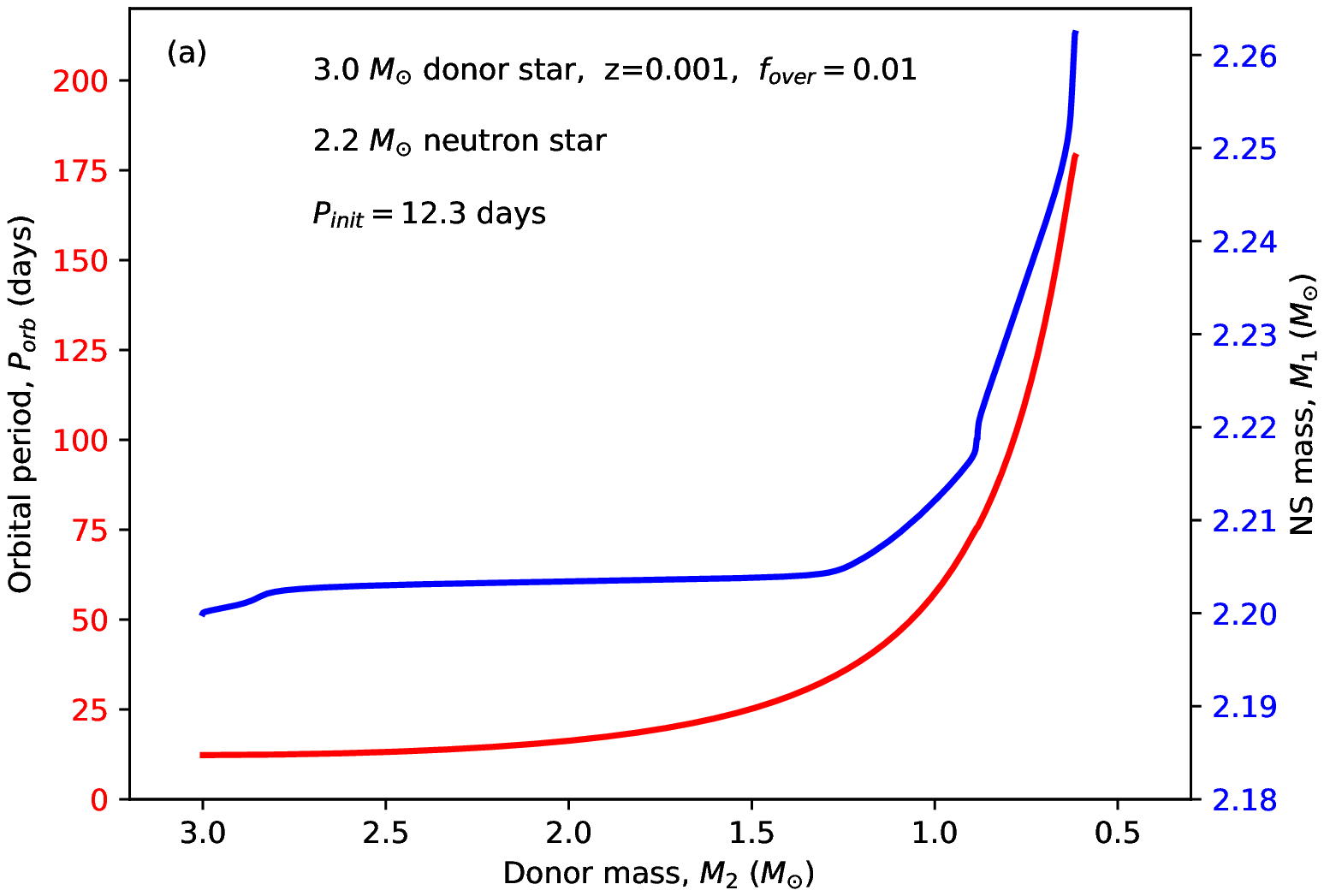}}
\subfigure{\includegraphics[scale=0.5]{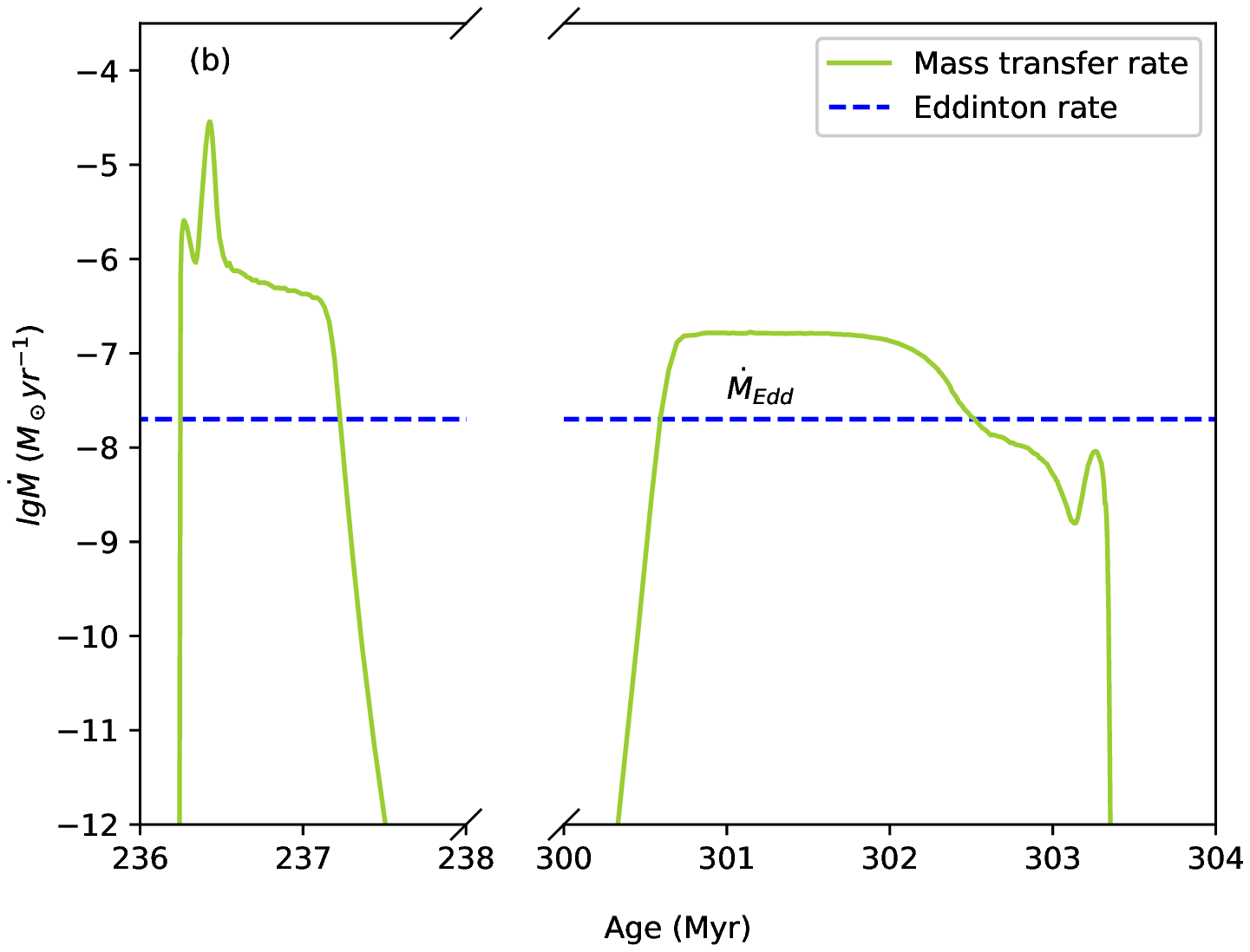}}
\subfigure{\includegraphics[scale=0.5]{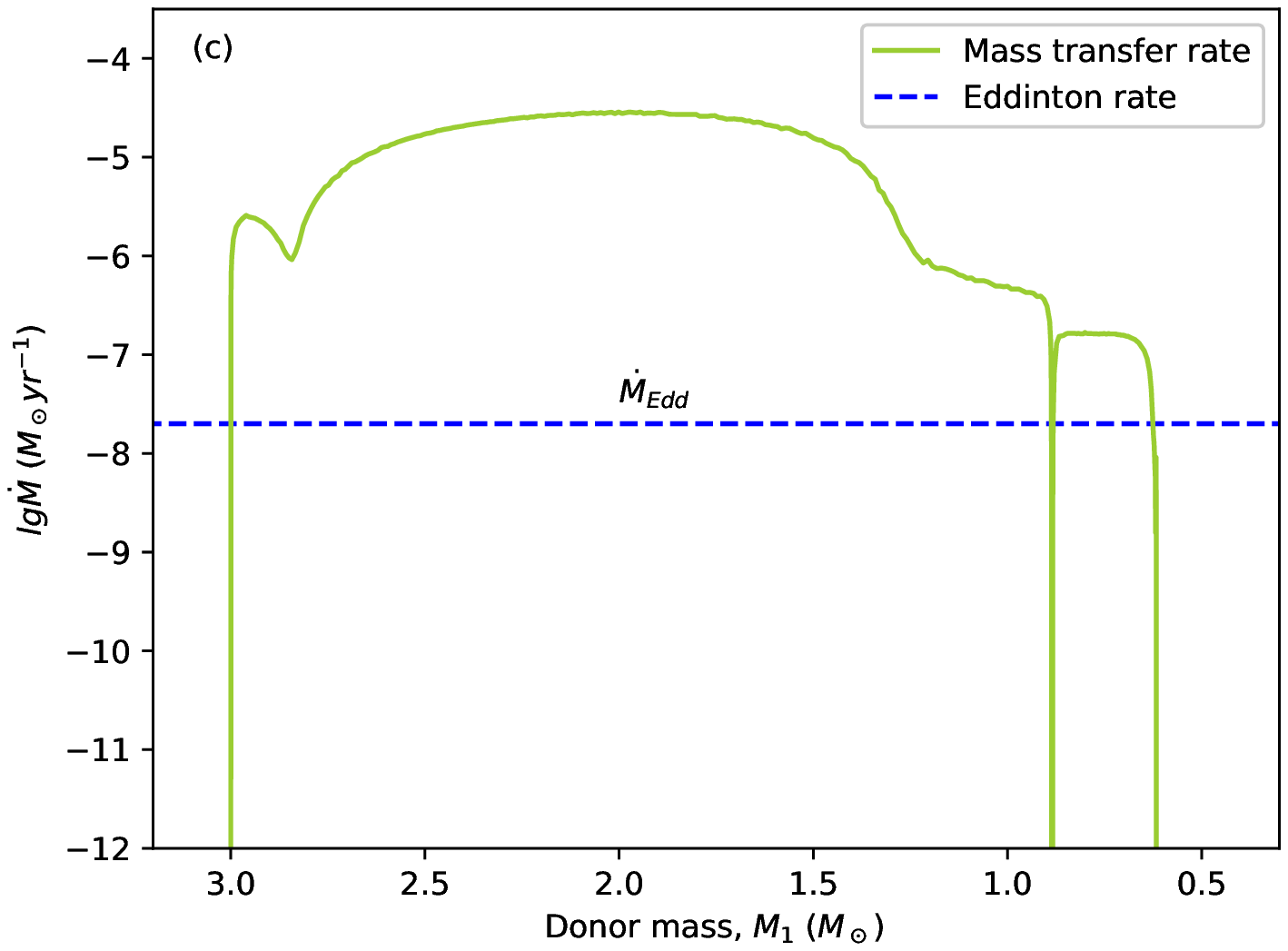}}
\subfigure{\includegraphics[scale=0.5]{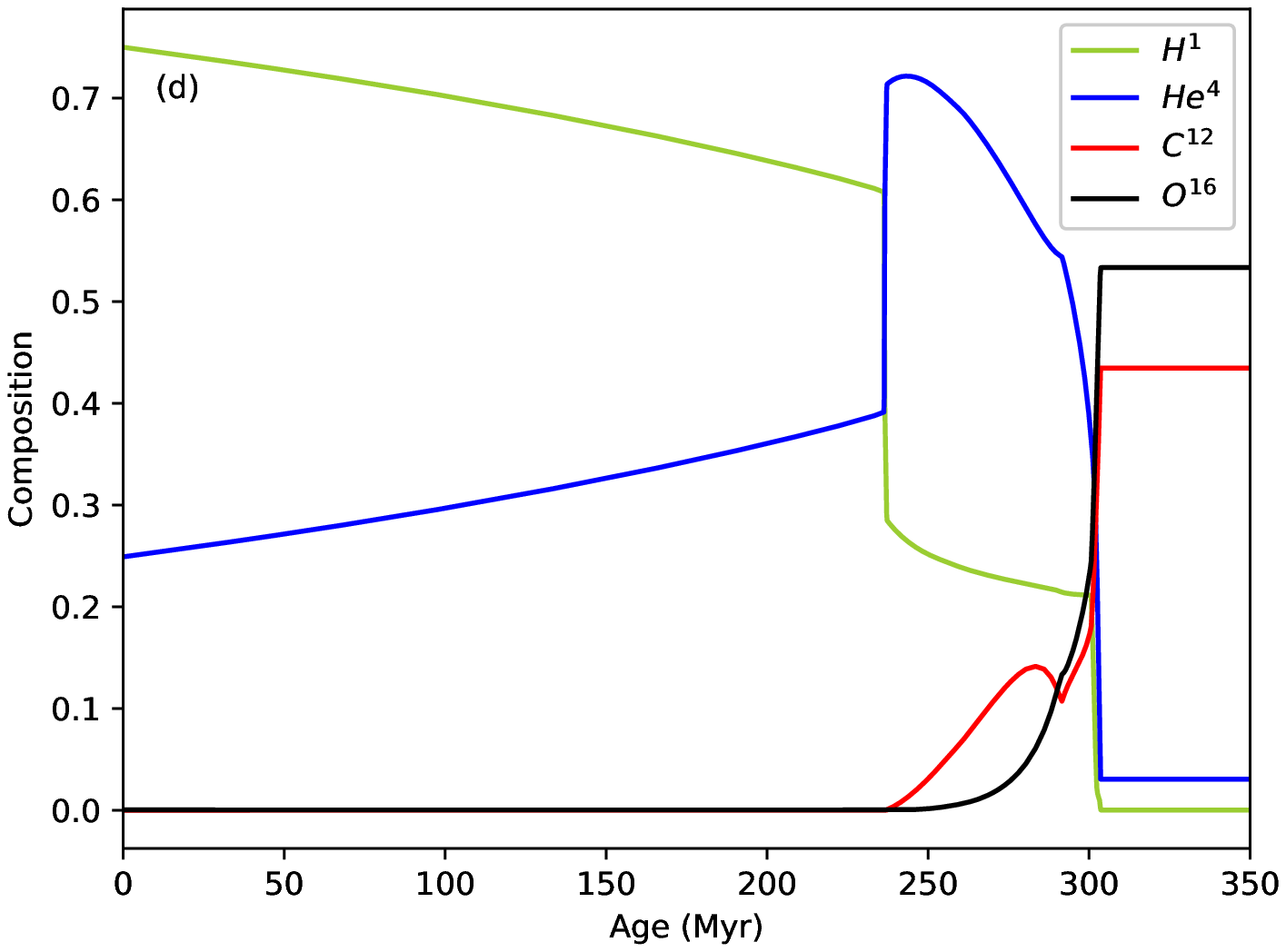}}
\caption{Evolution of an IMXB with the initial NS mass, donor mass and orbital period $M_{1}=2.2M_{\odot}$, $M_{2}=3.0M_{\odot}$, and $P_{\rm orb}=12.3$ d, respectively. The left panels show the evolution of $P_{\rm orb}$, $M_1$, and $\dot{M}$ as a function of the donor mass, and the right panels  the evolution of $\dot{M}$ and chemical composition for the donor star as a function of time.
\label{figure3}}

\end{figure}

Next we examine whether the mass transfer can accelerate the NS's spin period to milliseconds. The spin-up rate of the NS in a LMXB is determined by the rate of angular momentum transfer due to mass accretion,
\begin{equation}
2\pi I\dot{P}/P^2=\dot{M}_1(GM_1R_1)^{1/2},
\end{equation}
where $P$ and $\dot{P}$ are the spin period and its derivative of the NS respectively, and $R_1$ is the radius of the NS. Here we have assumed that the NS's magnetic field has been decayed so much that the accretion disk can extend to the surface of the NS. The amount of the accreted mass to produce a MSP can be roughly estimated to be
\begin{equation}
\Delta M\sim 0.1 M_\sun (M_1/2M_\sun)^{1/2}(R_1/10^6\,{\rm cm})^{-1/2}(P/3\,{\rm ms})^{-1}.
\end{equation}
The actual value of $\Delta M$ could be smaller by a factor of $\sim 2$ than that in Eq.~(13) if considering the effect of the NS magnetic field - accretion disk interaction. Combing this with Table 1, we notice that evolutions with Population II compositions seem to be more preferred for the formation of PSR J1640+2224.

Fig.~4 shows an example of the Population II evolutionary paths for PSR J1640+2224, which are able to reproduce the observed parameters of PSR J1640+2224. The evolution of this X-ray binary starts with a $2.2M_{\odot}$ NS and a $3.0M_{\odot}$ donor star in a 12.3 day orbit. The left panels show the evolution of $P_{\rm orb}$, $M_1$, and $\dot{M}_2$ as a function of the donor mass, and the right panels the evolution of $\dot{M}_2$ and the chemical composition of the donor star. At the age of $t=236$ Myr  the donor star overflows its RL and transfers mass to the NS at a rate well above the Eddington-limit accretion rate.  Accordingly, the mass transfer lasts $\sim$ 1.1 Myr, and the donor loses $\sim 3.1\,M_{\odot}$ mass. Due to the extensive mass loss, the orbital period does not decrease significantly. Then the donor star shrinks and becomes detached from its RL. The next mass transfer phase lasts $\sim$ 3.0 Myr with a  rate $\sim 10^{-7}-10^{-6}M_{\odot}$ yr$^{-1}$. After the mass transfer,  the NS has accreted material of $\sim 0.062M_{\odot}$.
The endpoint of the evolution is a binary consisting a recycled pulsar and a CO WD with a mass of $0.616M_{\odot}$.

\section{DISCUSSION AND CONCLUSIONS}

The observed properties of PSR J1640+2224, such as its short spin period (3.16 ms), high WD mass ($M_{2}\geq 0.4M_{\odot}$ with $>90\%$ confidence), long orbital period (175 days), and nearly circular orbit ($e=7.9725\times 10^{-4}$) make it distinct from other MSPs and imply something unusual in its evolutionary history.

In this work, we have attempted to explore which evolutionary channels can reproduce the observed parameters of PSR J1640+2224. To achieve this goal, we consider different NS mass, donor mass and metal abundance ($Z=0.02$ and 0.001). Our mainly results are presented in Fig.~3 and Tables 1-5, and can be summarized as follows.

1. For Population I chemical compositions, when $M_1\simeq 2.0 \,M_{\odot}$, it is possible to simultaneously account for the WD mass ($\sim 0.4 \,M_{\odot}$) and the orbital period ($\sim$ 175 days). But if the WD mass $> 0.6 \,M_{\odot}$, the NS mass should be larger than $2.2 \, M_{\odot}$, and the donor star must be initially of intermediate-mass.

2. For Population II chemical compositions, the evolution of original LMXBs containing a $1.4M_{\odot}$ NS and a $1.0M_{\odot}$ donor star may form PSR J1640+2224-like binaries with a $0.4M_{\odot}$ WD companion. However, if the WD indeed has a mass higher than $0.6M_{\odot}$, then the initial NS mass should be no less than $2.0M_{\odot}$, and the initial donor  mass should be higher than $3.0 \,M_{\odot}$.

3. When the NS spin evolution is taken into account, the evolutions with Population II compositions seem to be more preferred for the formation of PSR J1640+2224.

Almost all of our results predict a NS of initial mass higher than $2.0\,M_{\odot}$ (see Table 1). It is noted that \citet{Fo16} demonstrated that J1640+2224 is likely a massive NS ($4.4^{+2.9}_{-2.0}\,M_{\odot}$) by analyzing the North American Nanohertz Observatory for Gravitational Waves (NANOGrav) 9-year data set, although there are large uncertainties on the pulsar mass.

The birth masses of NSs depend on the physical mechanisms in core-collapse supernova explosions of massive stars, which are still highly uncertain \citep{T96,J12}. Theoretical studies do claim that NSs could be born massive. For example, by simulating neutrino-powered supernova explosions in spherical symmetry, \citet{U12} found that the birth masses of NSs vary within a range of $1.2-2.0M_{\odot}$. \citet{PT15} also numerically investigated the possible scope of the birth masses of NSs, and found that they could be as high as $1.9M_{\odot}$. \citet{S16} presented supernova simulations for a grid of massive stars and obtained the birth masses of NSs ranging from $1.2\,M_{\odot}$ to $1.8\,M_{\odot}$.

Observationally, super-massive NSs have been discovered in quite a few X-ray binaries and binary MSPs \citep{A16}. However, since they have experienced mass transfer processes, their current masses may not properly reflect their birth masses. For NSs in high-mass X-ray binaries (HMXBs), because they are relatively young (with an age $\lesssim 10^7$ yr) and the efficiency of wind accretion is very low, their masses should be very close to those at birth \citep{S12}. In HMXBs, the maximum NS mass measured is $2.12\pm 0.16\,M_{\odot}$ for Vela X-1 \citep{R11,F15}. For binary MSPs, the masses of super-massive NSs span between $\sim 2.0\,M_\sun$ and $\sim 2.9\,M_\sun$ \citep[][for a recent review]{L19}. For example, PSR J0740+6620 is one of the most massive NSs ever accurately measured, with a mass $2.14^{+0.10}_{-0.09}\,M_{\odot}$ \citep{C19}, a spin period of 2.88 ms and a He WD companion of mass $0.26M_{\odot}$. These parameters mean that PSR J0740+6620 should have experienced extensive mass accretion during the LMXB phase, so its current mass may not represent its birth mass. However, recent studies suggested that mass transfer in LMXBs is likely to be highly non-conservative even at sub-Eddington mass transfer rates \citep[e.g.,][]{T11,L11,A13,F16}.
Our results show that, even with Eddington-limited accretion, a super-massive newborn NS is still required, at least for PSR J1640+2224.

We need to mention that our results are subject to several uncertainties, such as the treatment of convection in the donor star and the accretion efficiency during the mass transfer process. The convective overshooting parameter plays an important role in our work. Up to the present time two standard methods are used in modelling overshooting in numerical calculation. The first one is based on a simple extension of the the convectively mixed region above the boundary defined by the Schwarzschild criterion. This extension $l_{\rm ov}$ is parameterized in terms of the local pressure scale height $H_{\rm P}$ at the boundary,
\begin{equation}
l_{\rm ov}=\alpha H_{\rm P},
\end{equation}
where $\alpha$ is the convective overshooting parameter. In this model, various attempts have been undertaken to constrain the value of $\alpha$. \citet{Sc97} derived it to be 0.25 and 0.32 for stars in the mass range of $(2.5-7) \,M_{\odot}$ for eclipsing binary stars. \citet{Ri00} and \citet{Cl07} found that the value of $\alpha$ is in the range of $0.1-0.6$, and the amount of overshooting increases systematically with the stellar mass. \citet{Sa12} used asteroseismic analysis and obtained $\alpha=0.1-0.3$ for $\beta$ Cephei stars (with mass $M \geq 8\,M_{\odot}$). \citet{De10} analyzed the period space of SPB star HD 50230 ($M=7-8M_{\odot}$) and suggested that the overshooting extent of the convective core is about $(0.2-0.3)\,H_{\rm P}$. \citet{P14} analyzed the period spacings of KIC 10526294 ($M=3.25M_{\odot}$) and suggested that $\alpha$ is less than or equal to 0.15.

In an alternative approach, convective overshooting is considered to be a diffusive process with a diffusion parameter
\begin{equation}
D=D_{0}\exp(\frac{-2z}{f_{\rm ov}H_{\rm P}})
\end{equation}
where $z$ is the radial distance from the formal Schwarzschild border and $f_{\rm ov}$ is a free parameter, $D_{0}$ is set as the scale of diffusive speed and derived from the convective velocity obtained from the mixing-length theory and taken below the Schwarzschild boundary \citep{H00}.

\citet{H00} investigated the evolution of AGB stars with convective overshooting and set the diffusive convective parameter $f_{\rm ov}=0.016$, which is widely adopted. \citet{Mo15} analyzed KIC 10526294 with both the step function overshooting and exponentially decreasing overshooting, and found that the latter is better than the former for interpreting the observations with $f_{\rm ov}=0.017-0.018$. \citet{Mo16} obtained $f_{\rm ov}=0.024\pm 0.001$ for KIC 7760680. Based on the $k-\omega$ model, \citet{Guo19} concluded that $f_{\rm ov}$ is about 0.008 for stars in the mass range of $(1.0-1.8)\,M_{\odot}$.


In Table 6, we compare the calculated results for I/LMXB evolution with different overshooting parameters ($\alpha=0.2$, $0.335$ and $f_{\rm ov}=0.016$). The initial NS and donor masses are taken to be $M_{1}=2.0\,M_{\odot}$ and $M_{2}=2.5\,M_{\odot}$, respectively. From the results in Table 6, we conclude that the overshooting model with $\alpha=0.2$ is nearly same as the diffusive overshooting model with $f_{\rm ov}=0.016$, and that increasing the value of $\alpha$ to 0.335 would result in more massive WDs and shorter orbital periods with the same initial parameters.

In Fig.~5, we compare the details of the evolution ($R_{2}, \dot{M}_{2}, P_{\rm orb}, M_{1}$ and $M_{2}$) for an IMXB with different overshooting parameters. The initial orbit period is set to be 5.35 days.  The evolutionary tracks with $\alpha=0.2$ and $f_{\rm ov}=0.016$ are nearly identical, while for $\alpha=0.335$ RLOF starts $\sim$ 60 Myr later than the others (panels (a) and (b)). In general, larger overshooting parameter produces a more massive core, less mass loss and smaller accreted mass by the NS (panel (d)).

In summary, PSR J1640+2224 was likely born massive, and this conclusion seems not sensitively dependent on the treatment of convective overshooting with reasonable parameters.

\begin{figure}[htbp]

\centerline{\includegraphics[scale=0.6]{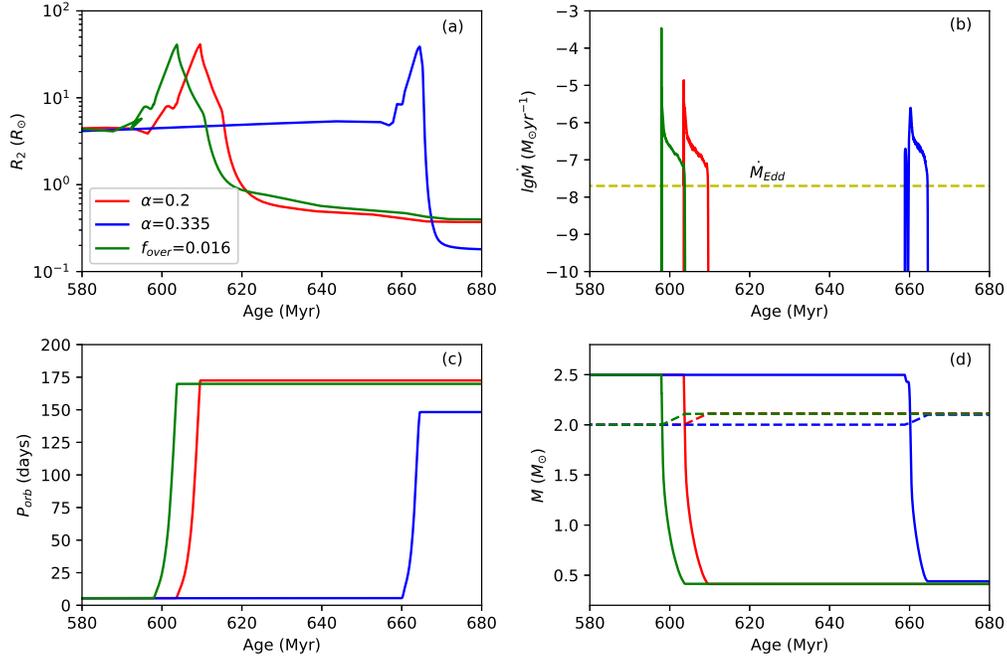}}
\caption{Comparison of the IMXB evolution with different overshoot parameters. The binary initially consists of a $2.0M_{\odot}$ NS and a $2.5M_{\odot}$ donor star, with an initial orbit period of 5.35 days. The top panels show the evolutions of $R_{2}$ and $\dot{M}_{2}$ as a function of time, and the bottom panels the evolutions of $P_{\rm orb}$ and stellar mass as a function of time. The red, blue and green curves represent the results with $\alpha =0.2$, $0.335$, and $f_{\rm ov}=0.016$, respectively. In panel (d), the solid and dashed lines denote the donor mass and the NS mass, respectively.
\label{figure7}}

\end{figure}

\acknowledgements
We are grateful the referee for helpful comments. This work was supported by the National Key Research and Development Program of China (2016YFA0400803), the Natural Science Foundation of China under grant No. 11773015 and Project U1838201 supported by NSFC and CAS.


\clearpage
\begin{table}
\footnotesize
\begin{center}
\caption{Calculated results of the binary evolution with $M_{1,\rm i}=1.4\,M_{\odot}$}        
\begin{tabular}{cccccc|cccccc}
  \hline
  \hline
  \multicolumn{6}{c}{Pop. \uppercase\expandafter{\romannumeral1};   $f_{\rm ov}=0.016$} & \multicolumn{6}{c}{Pop. \uppercase\expandafter{\romannumeral2};   $f_{\rm ov}=0.01$}\\
  \hline
  $M_{2,i}$ & $P_{\rm orb,i}$ & Case & $M_{\rm 1,f}$ & $M_{\rm 2,f}$ & $P_{\rm orb,f}$ & $M_{\rm 2,i}$ & $P_{\rm orb,i}$ & Case & $M_{\rm 1,f}$ & $M_{\rm 2,f}$ & $P_{\rm orb,f}$  \\
  ($M_{\odot}$) & (days) & (RLO) & ($M_{\odot}$) & ($M_{\odot}$) & (days) & ($M_{\odot}$) & (days) & (RLO) & ($M_{\odot}$) & ($M_{\odot}$) & (days)  \\
  \hline
  $1.0$  & 3.0 & B & 2.069 & 0.264 & 19.7 & $1.0$ & 3.0 & B & 2.037 & 0.300 & 25.8 \\
         & 5.0 & B & 2.040 & 0.294 & 52.9 &  & 5.0 & B & 2.015 & 0.324 & 43.4 \\
         & 10.0 & B & 1.994 & 0.316 & 97.5 &  & 10.0 & B & 1.966 & 0.350 & 75.9  \\
         & 20.0 & B & 1.890 & 0.338 & 172.0 &  & 20.0 & B & 1.866 & 0.379 & 129.4 \\
         & 60.0 & B & 16.22 & 0.382 & 436.1 &  & 60.0 & B & 1.576 & 0.437 & 317.0 \\
  $1.5$  & 3.0 & B & 2.389 & 0.281 &  34.9 & $1.5$  & 2.0 & B & 2.158 & 0.304 &  27.1  \\
         & 4.0 & B & 2.282 & 0.301 & 62.9 &   & 6.0 & B & 2.165 & 0.351 &  76.1  \\
         & 6.0 & B & 2.179 & 0.316 & 95.7 &   & 10.0 & B & 2.025 & 0.373 &  113.0 \\
         & 10.0& B & 2.089 & 0.333 & 147.1&   & 20.0 & B & 1.784 & 0.407 &  199.1 \\
         & 20.0& B & 1.827 & 0.358 & 267.1&   & 60.0 & B & 1.539 & 0.471 &  464.5 \\
  $2.0$    &1.5& A & 2.707 & 0.290 & 38.0 & $2.0$  & 1.0 & A & 1.635 & 0.324 &  21.3 \\
         & 2.0 & B & 2.058 & 0.305 & 51.1 &   & 2.0 & B & 1.604 & 0.337 &  39.1 \\
         &2.5& B & 1.886 & 0.317 & 58.8 &   & 3.0 & B & 1.619 & 0.346 &  54.1 \\
  $2.5$  &1.5& A & 2.469 & 0.275 & 18.0 &   & 5.0 & B & 1.649 & 0.374 &  74.6 \\
         & 2.0 & B & 1.624 & 0.345 & 32.7 & $2.5$  & 1.0 & A & 1.526 & 0.395 & 11.8 \\
         & 3.0 & B & 1.555 & 0.365 & 44.5 &   & 2.0 & B & 1.465 & 0.424 &  20.6  \\
         & 4.0 & B & 1.526 & 0.373 & 57.3 &   & 4.0 & B & 1.448 & 0.436 &  38.9 \\
  $3.0$    &1.5& A & 2.171 & 0.270 & 16.4 &   & 6.0 & B & 1.446 & 0.447 &  55.1 \\
         & 2.0 & A & 1.964 & 0.373 & 15.5 &   & 8.0 & B & 1.445 & 0.468 &  66.7 \\
         & 4.0 & B & 1.466 & 0.440 & 27.1 & $3.0$  & 1.0 & A & 1.481 & 0.462 &  6.1 \\
         & 6.0 & B & 1.456 & 0.447 & 39.6 &   & 3.0 & B & 1.428 & 0.534 &  13.5 \\
  $3.5$  &1.5& A & 2.015 & 0.253 & 11.1 &   & 5.0 & B & 1.421 & 0.540 &  26.4 \\
         & 2.0 & A & 1.786 & 0.379 & 9.8  &   & 8.0 & B & 1.419 & 0.541 &  35.0 \\
         & 3.0 & B & 1.444 & 0.511 & 9.2  &   & 12.0 & B & 1.417 & 0.547 &  51.3 \\
         & 5.0 & B & 1.434 & 0.527 & 14.4 & $3.5$  & 1.0 & A & 1.450 & 0.531 &  3.0 \\
         & 8.0 & B & 1.428 & 0.535 & 22.4 &   & 2.0 & B & 1.419 & 0.644 &  3.9 \\
  $4.0$    & 3.0 & B & 1.430 & 0.572 & 4.3 &   & 3.0 & B & 1.415 & 0.650 &  5.7  \\
         & 4.0 & B & 1.422 & 0.600 & 5.2 &   & 5.0 & B & 1.413 & 0.660 &  9.3  \\
         & 5.0 & B & 1.421 & 0.605 & 6.3 &   & 8.0 & B & 1.412 & 0.661 &  14.9 \\
         & 6.0 & B & 1.419 & 0.616 & 7.3 &   & 12.0 & B & 1.413 & 0.663 &  22.1  \\

    \bottomrule[1pt]
\end{tabular}
\end{center}
\end{table}

\begin{table}
\footnotesize
\begin{center}
\caption{Calculated results of the binary evolution with $M_{1,\rm i}=1.8M_{\odot}$}
\begin{tabular}{cccccc|cccccc}
  \hline
  \hline
  \multicolumn{6}{c}{Pop. \uppercase\expandafter{\romannumeral1};   $f_{\rm ov}=0.016$} & \multicolumn{6}{c}{Pop. \uppercase\expandafter{\romannumeral2};   $f_{\rm ov}=0.01$}\\
  \hline
  $M_{2,i}$ & $P_{\rm orb,i}$ & Case & $M_{\rm 1,f}$ & $M_{\rm 2,f}$ & $P_{\rm orb,f}$ & $M_{\rm 2,i}$ & $P_{\rm orb,i}$ & Case & $M_{\rm 1,f}$ & $M_{\rm 2,f}$ & $P_{\rm orb,f}$  \\
  ($M_{\odot}$) & (days) & (RLO) & ($M_{\odot}$) & ($M_{\odot}$) & (days) & ($M_{\odot}$) & (days) & (RLO) & ($M_{\odot}$) & ($M_{\odot}$) & (days) \\
  \hline
  $1.5$  & 3.0 & B & 2.877 & 0.295 &  53.4 & $1.5$  & 1.5 & A & 2.700 & 0.276 & 16.2\\
         & 4.0 & B & 2.816 & 0.312 & 85.1 & & 3.0 & B & 2.827 & 0.336 & 56.9 \\
         & 8.0 & B & 2.659 & 0.337 & 161.7 & & 10.0 & B & 2.536 & 0.387 & 145.8\\
         & 15.0 & B & 2.408 & 0.363 & 277.1 & & 20.0 & B & 2.228 & 0.425 & 258.7\\
  $2.0$    & 2.0 & A & 2.468 & 0.322  & 75.1 & & 40.0 & B & 2.025 & 0.466 & 434\\
         & 3.0 & B & 2.229 & 0.335 & 101.2 & $2.0$   & 1.0 & A & 2.131 & 0.363  & 36.3\\
         & 6.0 & B & 2.227 & 0.353 & 172.8 & & 3.0 & B & 2.080 & 0.378 & 77\\
         & 9.0 & B & 2.204 & 0.368 & 234.0 & & 10.0 & B & 2.066 & 0.475 & 143.2\\
         & 15.0 & B & 2.142 & 0.387 & 350.4 & & 20.0 & B & 1.937 & 0.492 & 283.1\\
  $2.5$  &1.5& A & 3.067 & 0.299 & 62.9 & & 60.0 & B & 1.833 & 0.549 & 663.9\\
         & 2.0 & A & 2.439 & 0.350 & 74.0 & $2.5$  & 1.0 & A & 1.944 & 0.397 & 26\\
         & 3.0 & B & 1.976 & 0.377 & 93.8 & & 3.0 & B & 1.862 & 0.446 & 59.8\\
         & 4.0 & B & 1.947 & 0.387 & 118.3 & & 5.0 & B & 1.852 & 0.502 & 74.9\\
  $3.0$    &1.5& A & 2.878 & 0.299 & 43.0 & & 10.0 & B & 1.885 & 0.530 & 126.0\\
         & 2.0 & A & 2.628 & 0.378 & 42.0 & $3.0$ & 1.0 & A & 1.890 & 0.468 & 16.1\\
         & 3.0 & B & 1.879 & 0.440 & 54.9 & & 3.0 & B & 1.830 & 0.537 & 35\\
         & 4.0 & B & 1.870 & 0.445 & 71.3 & & 5.0 & B & 1.823 & 0.547 & 56.1\\
         & 6.0 & B & 1.862 & 0.455 & 101.4 & & 10.0 & B & 1.837 & 0.598 & 107.4\\
  $3.5$  &1.5& A & 2.667 & 0.301 & 39.1 & $3.5$  & 1.0 & A & 1.985 & 0.524 & 9.4\\
         & 2.0 & A & 2.359 & 0.394 & 32.6 & & 3.0 & B & 1.824 & 0.660 & 17.7\\
         & 4.0 & B & 1.839 & 0.526 & 37.7 & & 5.0 & B & 1.817 & 0.666 & 28.8\\
         & 6.0 & B & 1.832 & 0.537 & 53.9 & & 10.0 & B & 1.817 & 0.670 & 56.1\\
         & 9.0 & B & 1.828 & 0.543 & 78.8 & & 15.0 & B & 1.816 & 0.672 & 32.6\\
  $4.0$    &1.5& A & 2.552 & 0.298 & 30.2 & $4.0$    & 1.0 & A & 2.013 & 0.556 & 6\\
         & 2.0 & A & 2.165 & 0.405 & 25.0 & & 3.0 & B & 1.816 & 0.746 & 10.2\\
         & 3.0 & B & 1.826 & 0.605 & 15.0 & & 5.0 & B & 1.813 & 0.762 & 16.2\\
         & 6.0 & B & 1.819 & 0.623 & 28.0 & & 12.0 & B & 1.811 & 0.776 & 38.0\\
         & 10.0& B & 1.815 & 0.643 & 43.4 & & 20.0 & B & 1.810 & 0.779 & 62.8\\
   \bottomrule[1pt]
\end{tabular}
\end{center}
\end{table}

\begin{table}
\footnotesize
\begin{center}
\caption{Calculated results of the binary evolution with $M_{1,\rm i}=2.0M_{\odot}$}
\begin{tabular}{cccccc|cccccc}
  \hline
  \hline
 \multicolumn{6}{c}{Pop. \uppercase\expandafter{\romannumeral1};   $f_{\rm ov}=0.016$} & \multicolumn{6}{c}{Pop. \uppercase\expandafter{\romannumeral2};   $f_{\rm ov}=0.01$}\\
  \hline
  $M_{2,i}$ & $P_{\rm orb,i}$ & Case & $M_{\rm 1,f}$ & $M_{\rm 2,f}$ & $P_{\rm orb,f}$ & $M_{\rm 2,i}$ & $P_{\rm orb,i}$ & Case & $M_{\rm 1,f}$ & $M_{\rm 2,f}$ & $P_{\rm orb,f}$  \\
  ($M_{\odot}$) & (days) & (RLO) & ($M_{\odot}$) & ($M_{\odot}$) & (days) & ($M_{\odot}$) & (days) & (RLO) & ($M_{\odot}$) & ($M_{\odot}$) & (days) \\
  \hline
  $1.5$  & 3.0 & B & 3.082 & 0.300 & 61.3 & $1.5$  & 2.0 & B & 2.861 & 0.322 & 42.2\\
         & 4.0 & B & 3.049 & 0.315 & 94.3 & & 5.0 & B & 2.977 & 0.361 & 92.8 \\
         & 6.0 & B & 2.975 & 0.330 & 139.2 & & 10.0 & B & 2.769 & 0.393 & 158.7 \\
         & 10.0& B & 2.863 & 0.348 & 212.6 & & 20.0 & B & 2.44 & 0.431 & 281.4 \\
         & 20.0& B & 2.493 & 0.378 & 391.6 & & 40.0 & B & 2.233 & 0.473 & 475 \\
  $2.0$    & 2.0 & A & 2.681 & 0.325 & 87.5 &$2.0$& 1.0 & A & 2.365 & 0.331 & 43.0\\
         & 3.0 & B & 2.464 & 0.340 & 117.9 &  & 3.0 & B & 2.337 & 0.390 & 81.4\\
         & 6.0 & B & 2.459 & 0.360 & 199.1 &  & 6.0 & B & 2.318 & 0.477 & 99.8\\
         & 9.0 & B & 2.429 & 0.374 & 272.2 &  & 10.0 & B & 2.283 & 0.487 & 162.9\\
         & 15.0& B & 2.355 & 0.395 & 406.8 &  & 15.0 & B & 2.205 & 0.512 & 225.4\\
  $2.5$  &1.5& A & 3.326 & 0.309 & 78.9 &$2.5$&1.0& A & 2.149 & 0.397 & 33.8\\
         & 2.0 & A & 2.705 & 0.350 & 94.1 &   & 3.0 & B & 2.068 & 0.469 & 69.0\\
         & 3.0 & B & 2.183 & 0.380 & 119.8 &  & 8.0 & B & 2.105 & 0.528 & 131.2\\
         & 5.0 & B & 2.114 & 0.407 & 166.2 &  &16.0 & B & 2.095 & 0.538 & 253.1\\
  $3.0$    &1.5& A & 3.138 & 0.309 & 64.9 & $3.0$ &1.0& A & 2.094 & 0.467 & 22.5\\
         & 2.0 & A & 2.847 & 0.382 & 61.1 &   & 3.0 & B & 2.030 & 0.527 & 51.1\\
         & 3.0 & B & 2.091 & 0.437 & 78.4 &   & 8.0 & B & 2.050 & 0.597 & 98.6\\
         & 7.0 & B & 2.055 & 0.479 & 148.7 & & 15.0 & B & 2.056 & 0.624 & 163.3\\
  $3.5$  &1.5& A & 2.928 & 0.315 & 56.9 &$3.5$&1.0& A & 2.200 & 0.525 & 14.2 \\
         & 2.0 & A & 2.634 & 0.397 & 48.7 &   & 2.0 & B & 2.024 & 0.614 & 20.7\\
         & 3.0 & B & 2.051 & 0.509 & 45.9 &   & 6.0 & B & 2.013 & 0.644 & 55.6\\
         & 6.0 & B & 2.032 & 0.540 & 79.2  &  & 10.0& B & 2.013 & 0.652 & 89.9\\
         & 10.0& B & 2.027 & 0.549 & 127.1 &  & 18.0& B & 2.032 & 0.711 & 130.4\\
  $4.0$    &1.5& A & 2.761 & 0.316 & 49.1 &$4.0$&1.0  & A & 2.231 & 0.562 & 9.6 \\
         & 2.0 & A & 2.404 & 0.408 & 40.6 &   & 2.0 & B & 2.022 & 0.705 & 12.2 \\
         & 3.0 & B & 2.030 & 0.591 & 25.6 &   & 6.0 & B & 2.014 & 0.740 & 32.9 \\
         & 6.0 & B & 2.020 & 0.626 & 44.1  &  & 12.0& B & 2.012 & 0.750 & 64.0 \\
         & 12.0& B & 2.017 & 0.625 & 88.9  &  & 20.0& B & 2.011 & 0.756 & 104.5 \\
   \bottomrule[1pt]
\end{tabular}
\end{center}
\end{table}

\begin{table}
\footnotesize
\begin{center}
\caption{Calculated results of the binary evolution with $M_{1,\rm i}=2.2M_{\odot}$}
\begin{tabular}{cccccc|cccccc}
  \hline
  \hline
  \multicolumn{6}{c}{Pop. \uppercase\expandafter{\romannumeral1};   $f_{\rm ov}=0.016$} & \multicolumn{6}{c}{Pop. \uppercase\expandafter{\romannumeral2};   $f_{\rm ov}=0.01$}\\
  \hline
  $M_{2,i}$ & $P_{\rm orb,i}$ & Case & $M_{\rm 1,f}$ & $M_{\rm 2,f}$ & $P_{\rm orb,f}$ & $M_{\rm 2,i}$ & $P_{\rm orb,i}$ & Case & $M_{\rm 1,f}$ & $M_{\rm 2,f}$ & $P_{\rm orb,f}$  \\
  ($M_{\odot}$) & (days) & (RLO) & ($M_{\odot}$) & ($M_{\odot}$) & (days) & ($M_{\odot}$) & (days) & (RLO) & ($M_{\odot}$) & ($M_{\odot}$) & (days) \\
  \hline
  $1.5$  & 3.0 & B & 3.285 & 0.304 & 68.5 & $1.5$  & 1.5 & A & 3.079 & 0.298 & 24.5\\
         & 5.0 & B & 3.239 & 0.327 & 127.9 & & 5.0 & B & 3.203 & 0.365 & 99.5 \\
         & 10.0 & B & 3.093 & 0.352 & 228.8 & & 10.0 & B & 2.993 & 0.397 & 170.1 \\
         & 20.0& B & 2.704 & 0.382 & 422.1 & & 20.0 & B & 2.650 & 0.436 & 300.8 \\
         & 40.0& B & 2.483 & 0.415 & 729.1 & & 40.0 & B & 2.439 & 0.479 & 504.2 \\
  $2.0$    & 1.5 & A & 3.563 & 0.314 & 73.1 &$2.0$& 1.0 & A & 2.592 & 0.334 & 49.1\\
         & 4.0 & B & 2.694 & 0.351 & 166.4 &  & 5.0 & B & 2.525 & 0.472 & 99.4\\
         & 8.0 & B & 2.658 & 0.376 & 280.3 &  & 10.0 & B & 2.477 & 0.480 & 198.4\\
         & 12.0 & B & 2.599 & 0.392 & 384.5 &  & 20.0 & B & 2.365 & 0.506 & 364.1\\
         & 15.0& B & 2.565 & 0.401 & 456.8 &  & 60.0 & B & 2.243 & 0.554 & 894.0\\
  $2.5$  &1.5& A & 3.572 & 0.320 & 89.3 &$2.5$&1.0& A & 2.362 & 0.400 & 41.1\\
         & 4.0 & B & 2.344 & 0.409 & 163.9 &   & 3.0 & B & 2.264 & 0.489 & 76.5\\
         & 8.0 & B & 2.313 & 0.490 & 196.5 &  & 8.0 & B & 2.300 & 0.533 & 157.5\\
         & 12.0 & B & 2.312 & 0.494 & 288.1 &  &20.0 & B & 2.095 & 0.552 & 364.6\\
         & 17.0 & B & 2.303 & 0.500 & 396.1 &  &40.0 & B & 2.095 & 0.661 & 463\\
  $3.0$    &1.5& A & 3.339 & 0.321 & 91.7 & $3.0$ &1.0& A & 2.295 & 0.469 & 29.1\\
         & 2.0 & A & 2.998 & 0.386 & 82.9 &   & 3.0 & B & 2.231 & 0.542 & 61.8\\
         & 4.0 & B & 2.276 & 0.457 & 121.5 &   & 8.0 & B & 2.251 & 0.608 & 121.2\\
         & 8.0 & B & 2.244 & 0.516 & 187.8 & & 15.0 & B & 2.249 & 0.633 & 204.6\\
  $3.5$  &1.5& A & 3.168 & 0.325 & 76.5 &$3.5$&1.0& A & 2.471 & 0.520 & 19.5 \\
         & 3.0 & A & 2.252 & 0.511 & 62.7 &   & 2.0 & B & 2.223 & 0.653 & 24.2\\
         & 5.0 & B & 2.235 & 0.538 & 91.5 &   & 6.0 & B & 2.218 & 0.667 & 69.5\\
         & 8.0 & B & 2.230 & 0.551 & 138.2  &  & 10.0& B & 2.218 & 0.696 & 105.7\\
         & 11.5& B & 2.225 & 0.582 & 168.2 &  & 20.0& B & 2.240 & 0.752 & 169.4\\
  $4.0$    &1.5& A & 2.998 & 0.330 & 69.3 &$4.0$&1.0  & A & 2.452 & 0.562 & 14.1 \\
         & 3.0 & A & 2.225 & 0.617 & 33.3 &   & 2.0 & A & 2.222 & 0.731 & 16.2 \\
         & 6.0 & B & 2.220 & 0.628 & 63.5 &   & 6.0 & B & 2.212 & 0.775 & 43.3 \\
         & 9.0 & B & 2.216 & 0.643 & 89.8  &  & 10.0& B & 2.212 & 0.778 & 71.4 \\
         & 15.0& B & 2.216 & 0.632 & 156.8  &  & 20.0& B & 2.240 & 0.787 & 138.5 \\
   \bottomrule[1pt]
\end{tabular}
\end{center}
\end{table}

\begin{table}
\begin{center}
\caption{IMXB evolution with different overshoot parameters}
\begin{tabular}{ccccccc}
  \toprule[2pt]
  \multicolumn{7}{c}{$M_{1}=2.0\,M_{\odot}$, $M_{2}=2.5\,M_{\odot}$} \\
  \hline
    & \multicolumn{2}{c}{$f_{\rm ov}=0.016$} & \multicolumn{2}{c}{$\alpha =0.2$} & \multicolumn{2}{c}{$\alpha =0.335$} \\
  \hline
  $P_{\rm orb, i}$  & $M^{\rm f}_{2}$ & $P_{\rm orb, f}$ & $M^{\rm f}_{2}$ & $P_{\rm orb, f}$ & $M^{\rm f}_{2}$ & $P_{\rm orb, f}$  \\
  (d)  & ($M_{\odot}$)& (d) & ($M_{\odot}$) & (d) & ($M_{\odot}$) & (d)  \\

  \hline
  $1.5$   & 0.309 & 78.9  & 0.307 & 80.4  & 0.309 & 67.8     \\
  $2.0$   & 0.350 & 94.1  & 0.352 & 90.7  & 0.351 & 85.3     \\
  $2.5$   & 0.368 & 106.7  & 0.373 & 103.0  & 0.389 & 86.5     \\
  $3.0$   & 0.380 & 119.8  & 0.379 & 120.3  & 0.411 & 91.8     \\
  $3.5$   & 0.387 & 134.0  & 0.383 & 137.5  & 0.414 & 104.3     \\
  $4.0$   & 0.394 & 146.2  & 0.389 & 151.0  & 0.419 & 117.0    \\
  $5.0$   & 0.407 & 166.2  & 0.405 & 168.1  & 0.428 & 141.7     \\
  $5.35$   & 0.414 & 170.0 & 0.411 & 172.5  & 0.439 & 148.2     \\
  $6.0$   & -- & --  & -- & --  & 0.443 & 160.9 \\
  $7.0$   & -- & -- & -- & --  & 0.451 & 179.8 \\
  $8.0$   & -- &  -- & -- & --  & 0.461 & 193.7 \\
  $10.0$  & -- & --  & -- & -- & -- & --\\
  \bottomrule[2pt]
\end{tabular}
\end{center}
\end{table}

\end{document}